\documentclass[12pt,preprint]{aastex}
\shortauthors{Marois et al.}
\begin{document}

\title{Confidence Level and Sensitivity Limits in High Contrast Imaging\altaffilmark{1}} 

\author{Christian Marois\altaffilmark{2}, David Lafreni\`{e}re\altaffilmark{3}, Bruce Macintosh\altaffilmark{3}, \& Ren\'{e} Doyon\altaffilmark{c}}

\email{cmarois@igpp.ucllnl.org, david@astro.umontreal.ca, bmac@igpp.ucllnl.org, doyon@astro.umontreal.ca}

\altaffiltext{1}{Based on observations obtained at the Canada-France-Hawaii Telescope (CFHT) which is operated by the National Research Council of Canada, the Institut National des Sciences de l'Univers of the Centre National de la Recherche Scientifique of France, and the University of Hawaii. Based on observations obtained at the Gemini Observatory, which is operated by the Association of Universities for Research in Astronomy, Inc., under a cooperative agreement with the NSF on behalf of the Gemini partnership: the National Science Foundation (United States), the Particle Physics and Astronomy Research Council (United Kingdom), the National Research Council (Canada), CONICYT (Chile), the Australian Research Council (Australia), CNPq (Brazil) and CONICET (Argentina).}

\altaffiltext{2}{Institute of Geophysics and Planetary Physics L-413,\\ Lawrence Livermore National Laboratory, 7000 East Ave, Livermore, CA 94550} 
\altaffiltext{3}{D\'{e}partement de physique and Observatoire du Mont M\'{e}gantic, Universit\'{e} de Montr\'{e}al, C.P. 6128, Succ. Centre-Ville,\\ Montr\'{e}al, QC, Canada H3C 3J7}

\begin{abstract} 
In long adaptive optics corrected exposures, exoplanet detections are currently limited by speckle noise originating from the telescope and instrument optics, and it is expected that such noise will also limit future high-contrast imaging instruments for both ground and space-based telescopes. Previous theoretical analysis have shown that the time intensity variations of a single speckle follows a modified Rician. It is first demonstrated here that for a circular pupil this temporal intensity distribution also represents the speckle spatial intensity distribution at a fix separation from the point spread function center; this fact is demonstrated using numerical simulations for coronagraphic and non-coronagraphic data. The real statistical distribution of the noise needs to be taken into account explicitly when selecting a detection threshold appropriate for some desired confidence level. In this paper, a technique is described to obtain the pixel intensity distribution of an image and its corresponding confidence level as a function of the detection threshold. Using numerical simulations, it is shown that in the presence of speckles noise, a detection threshold up to three times higher is required to obtain a confidence level equivalent to that at $5\sigma$ for Gaussian noise. The technique is then tested using TRIDENT CFHT and angular differential imaging NIRI Gemini adaptive optics data. It is found that the angular differential imaging technique produces quasi-Gaussian residuals, a remarkable result compared to classical adaptive optic imaging. A power-law is finally derived to predict the $1-3\times 10^{-7}$ confidence level detection threshold when averaging a partially correlated non-Gaussian noise.
\end{abstract}


\section{Introduction}
Searching for faint point sources around bright objects is a challenging endeavor. The atmosphere \citep{roddier1981,racine1999,macintosh2005}, telescope and instruments optics \citep{marois2003,marois2005,hinkley2007} produce speckles having a range of timescales that limit the direct detection of faint companions. From previous theoretical analysis, it is known that the speckle intensity temporal distribution is a modified Rician \citep{Goodman1968,soummer2004,fitzgerald2006}. If a large number of uncorrelated speckle realizations are coadded, from the central limit theorem, then the final residual speckle noise follows a Gaussian intensity distribution. Since atmospheric turbulence produces random speckles that have a very short correlation time, a Gaussian distributed residual speckle noise is commonly assumed for ground-based adaptive optics (AO) long integrations and a detection threshold of 5$\sigma$ is usually considered.

\begin{figure}
\epsscale{1}
\plotone{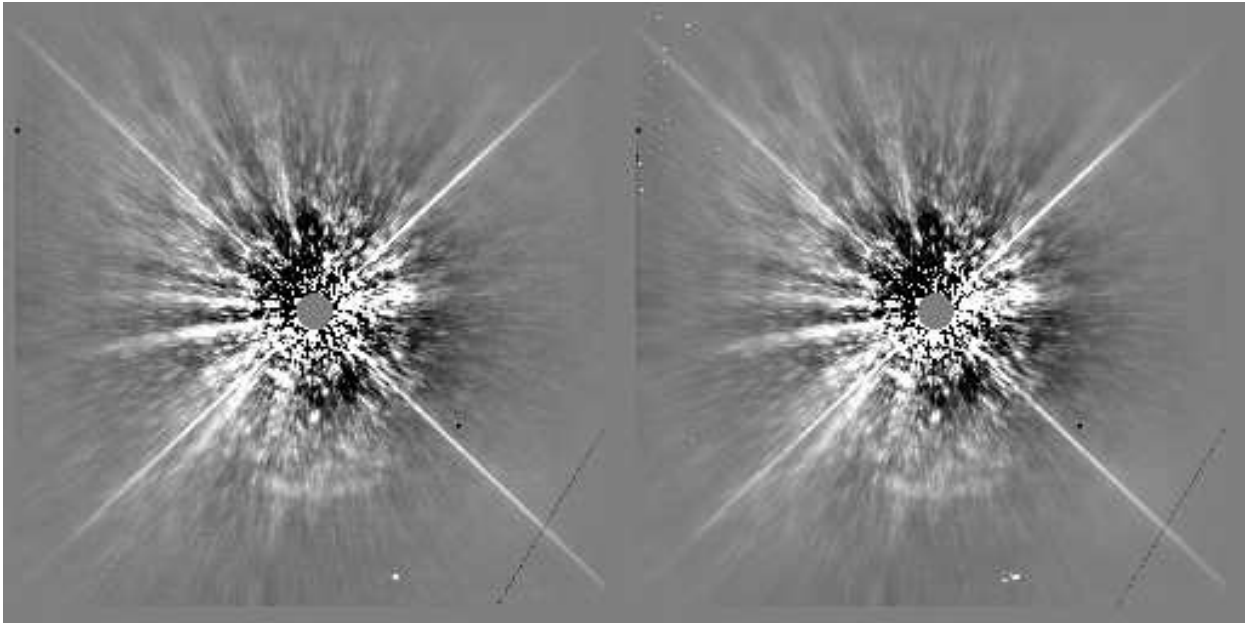}
\caption{Gemini NIRI/Altair quasi-static PSF of the star HD97334B. A single 30s exposure is shown (left) as well as the median of 90 30s exposures (right). The inner saturated region has been masked. A symmetric radial profile has been subtracted to highlight the speckle noise. FOV is $20^{\prime \prime} \times 20^{\prime \prime}$. Both images have the same intensity range and linear gray scale.\label{f1}}
\end{figure}

However, careful residual noise analysis of AO images have demonstrated that long exposures are not limited by random short-lived atmospheric speckles but by quasi-static speckles \citep{marois2003,marois2004phd,masciadri2005,marois2005,marois2006} originating from the telescope and instruments. The speckle noise currently limiting high-contrast ground-based imaging is thus very similar to that limiting space-based observations \citep{schneider2003}. The typical lifetime of ground-based quasi-static speckles has been found to be several minutes to hours \citep{marois2006,hinkley2007}; the noise in the combination of several images spanning $\sim$1~hr is very similar to that in a single image (see Fig.~\ref{f1} for an example; acquired with NIRI/Altair at the Gemini telescope). In this case, since the quasi-static speckle noise is well correlated for the entire sequence, the central limit theorem does not apply and the speckle noise in the final combined image will be non-Gaussian. Sensitivity limits calculated assuming Gaussian statistics would have lower confidence levels (CL). Finding a robust technique to estimate proper sensitivity limits is fundamental to analyze adequately the sensitivity of an exoplanet survey as a function of angular separation. The contrast limit reached by a survey plays a central role in Monte Carlo simulations to derive exoplanet frequencies around stars and constrain planet formation scenarios \citep{metchev2006,carson2006,kasper2007,lafreniere2007a}. Understanding the residual noise statistical distribution is thus important for future dedicated surveys of next generation AO systems like NICI \citep{ftaclas2003}, the Gemini Planet Imager (GPI, \citealp{macintosh2006}), the VLT SPHERE \citep{dohlen2006}, and as well as future space observatories.

In this paper, a new technique is presented to estimate sensitivity limits of a noise showing arbitrary statistics using a CL approach. The theory behind speckle statistics is summarized in \S~\ref{theo}. Then \S~\ref{ci} presents a technique to derive detection thresholds using the probability density function and associated CLs. The technique is applied to simulated (\S~\ref{app}) and observational (\S~\ref{obsdata}) data to confirm the theory and to validate the technique. The effect of averaging a sequence of independent non-Gaussian noise realizations is discussed in \S~\ref{dis}. Concluding remarks follow in \S~\ref{con}.

\section{Speckle Noise Statistics \label{theo}}
\begin{figure}
\epsscale{1}
\plotone{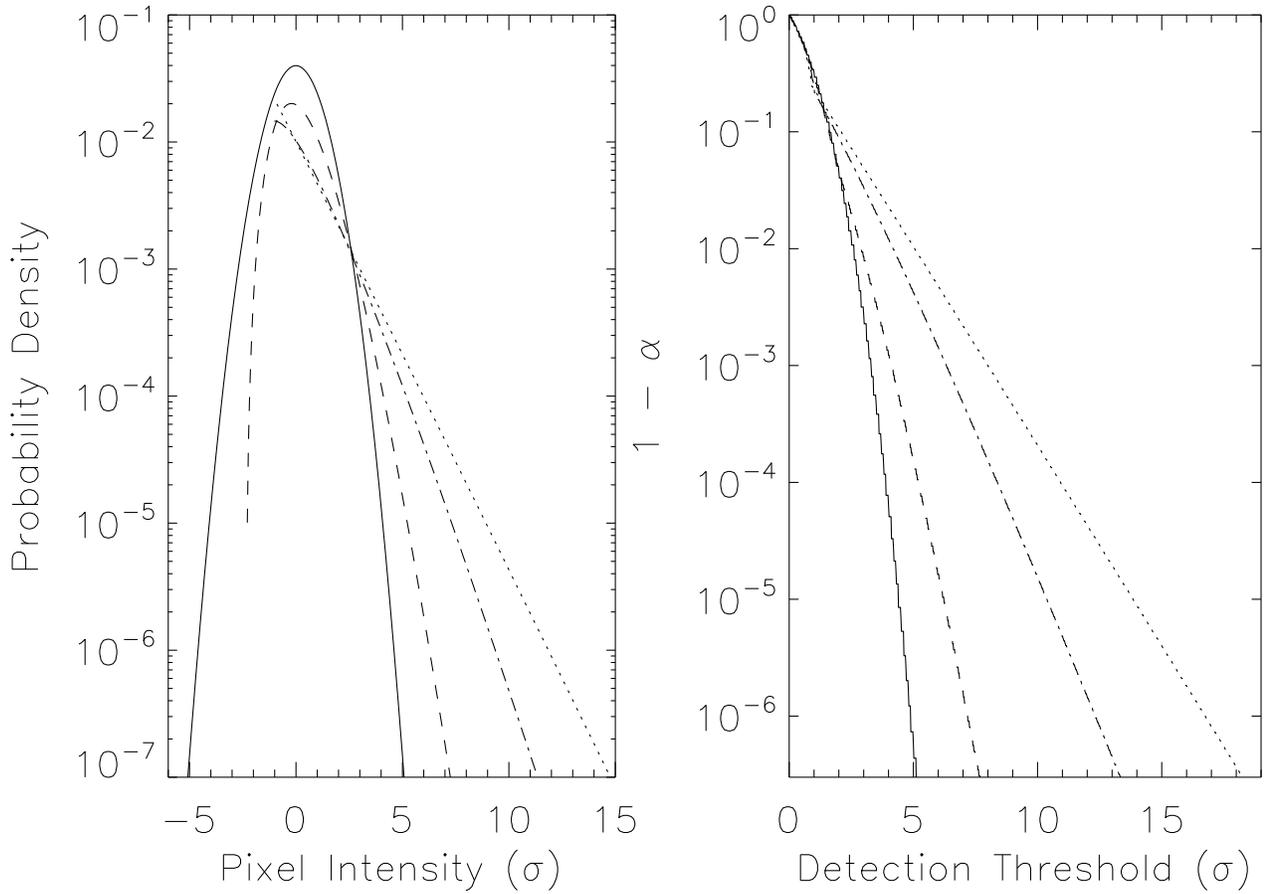}
\caption{Left panel: PDF for a Gaussian distribution (solid line) and modified Rician with $I_c/I_s = 10$ (dashed line), $I_c/I_s = 1$ (dot-dashed line), and $I_c/I_s = 0.1$ (dotted line). Right: corresponding CL as a function of detection threshold.\label{f2}}
\end{figure}

Following the work of \citet{Goodman1968,soummer2004,fitzgerald2006}, the speckle intensity probability density function (PDF) for one location in the image plane and random temporal phase errors can be shown to be a modified Rician (MR) function. At a specific location in the image plane, the MR PDF $p_{\rm{MR}}(I)$ is a function of the local time-averaged static point spread function intensity $I_c$ and random speckle noise intensities $I_s$:
\begin{equation}
p_{\rm{MR}}(I) = \frac{1}{I_s} \exp \left( - \frac{I+I_c}{I_s} \right) I_0 \left( \frac{2\sqrt{I I_c}}{I_s} \right)\rm{,}
\end{equation}
\noindent where $I$ is the point spread function (PSF) intensity ($I=I_c + I_s$) and $I_0 (x)$ is the zero-order modified Bessel function of the first kind. At a specific point of the PSF, if $I_c \gg I_s$, relevant to Airy ring pinned speckles, the associated PDF is a Gaussian-like function showing a bright positive tail, while if $I_c \ll I_s$, relevant to PSF dark rings or coronagaphic PSFs dominated by second order halo speckles, the noise distribution is exponential. The CL $\alpha$ for a given detection threshold $d$ is simply obtained by:
\begin{equation}
\alpha (d) = \int_{-d}^{d} p_{\rm{{MR}}}^{\prime} (I) dI\rm{,}\label{eqCL}
\end{equation}
\noindent where $p_{\rm{MR}}^{\prime}$ is the mean-subtracted PDF. Fig.~\ref{f2} illustrates the different possible regimes compared to a Gaussian intensity distribution. For a 5$\sigma$ detection threshold, where $\sigma$ is the standard deviation of the noise obtained using the {\em robust\_sigma} IDL algorithm\footnote{The {\em robust\_sigma} algorithm uses the median absolute deviation as a first estimate of the standard deviation and then weight points using Tukey's Biweight; this algorithm provides another step of robustness to avoid biasing the standard deviation estimate if bad pixels are present.}, a Gaussian distribution shows a $1-3\times 10^{-7}$ CL while a MR distribution show $\sim 1-10^{-2}$ to $1-10^{-3}$ CL. The MR distribution is thus producing much more false positive events. For example, consider a survey of many stars where each observation has a $500\times 500 \lambda/D$ field of view (FOV, i.e. the $20^{\prime \prime}\times 20^{\prime \prime}$ NIRI/Gemini FOV at H-band). If a $5\sigma$ detection threshold is selected, the Gaussian noise distribution would lead to one false positive detection every four stars while the MR distribution would lead to $\sim $250 to $\sim $2,500 false positives per star. A detection threshold two to three times higher is required for the MR distribution to show the same CL as a 5$\sigma$ Gaussian noise and the same number of false positive events.

In the previous speckle PDF analysis, it was shown that the atmospheric speckle noise PDF is obtained by analyzing the temporal variation at one location of the PSF. For a quasi-static speckle noise, this approach is not adequate since the noise does not vary significantly with time. The quasi-static noise PDF can be derived using a very simple argument. If we consider a PSF produced by a circular aperture and if the PDF is obtained by analyzing pixels inside a narrow annulus centered on the PSF core, azimuthal quasi-static speckle noise variations $I_s$ are produced with the same value of $I_c$ (here, $I_c$ is the unaberrated PSF and it is azimuthally symmetric for a circular aperture). The speckle noise inside a narrow annulus and from a single speckle noise realization thus shows the same PDF as a temporal speckle noise variation from random phase screens at any location inside the annulus.

\section{Experimental Derivation of the PDF and CL Curves\label{ci}}
A robust technique to derive sensitivity limits can be developed using CLs. The pixel PDF inside a specific region of the image is first obtained and the CL curve is then derived and extrapolated to estimate a local detection threshold. To avoid having too many false positive detections without missing possible faint companions, a CL of $1-3\times 10^{-7}$ (5$\sigma$ if Gaussian) is selected here. The basic steps to derive the PDF, to obtain the CL curves and to estimate the $1-3\times 10^{-7}$ CL detection threshold is summarized in table~\ref{tabpdfstep}.

The local PDF is obtained by producing an histogram of the pixel intensities inside a specific region of the image after subtraction of the mean pixel intensity over the region and division by the noise RMS of the image region. The CL curve as a function of detection threshold can be easily estimated by integrating the PDF inside the interval $\pm d$ (see Eq.~\ref{eqCL}). Due to the limited number of resolution elements (one $\lambda/D$ for PSFs or a pixel for simulated noise images) in an image, the PDF and CL curves will be known up to a certain detection threshold. In theory, for the $1-3\times 10^{-7}$ CL detection threshold considered here, each area where the PDF needs to be estimated should have several million independent resolution elements. In practice, for images typically containing up to $500\times 500 \lambda/D$ (250,000 resolution elements), the PDF will be known only up to $\sim 1 - 10^{-5}$ CL for Gaussian noise. A model fit using a $\chi^2$ analysis or a polynomial fit are required to extrapolate the CL curve and obtain the detection threshold corresponding to a $1-3\times 10^{-7}$ CL. Since the CL curves of various distributions are nearly linear in a semi-$\log (1-\alpha)$ vs detection threshold plot (see Fig.~\ref{f2}), we have chosen to use a polynomial fit due to its simplicity of implementation, its execution speed, and its accuracy. Due to non-linear effects for detection thresholds near 0$\sigma$, a linear fit is first performed for detection thresholds above 1.5$\sigma$. If the detection threshold for a $1-3\times 10^{-7}$ CL is below 9$\sigma$, a second order polynomial fit is used instead to better approximate the CL curve for quasi-Gaussian statistics. The CL extrapolation accuracy will be analyzed in the next section.

\section{Technique Validation with Simulated Data\label{app}}
In this section, the PDF, the CL curve and the $1-3\times 10^{-7}$ detection threshold of simulated data are obtained.

\subsection{Simulated PDFs}
\begin{figure}
\epsscale{1}
\plotone{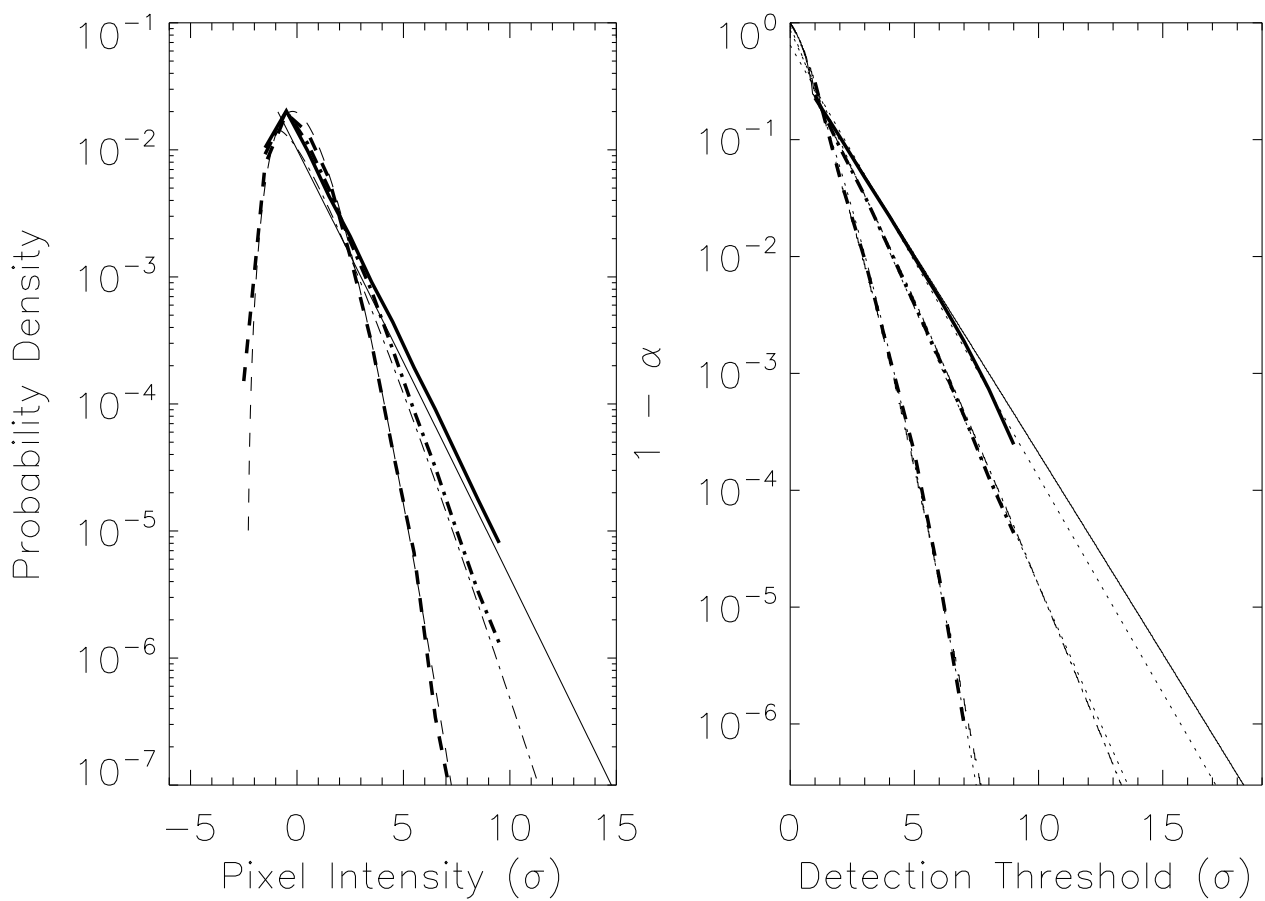}
\caption{The PDF (left panel) and CL curves (right panel) obtained from a simulated noise image ($10^{6}$ resolution elements) generated with an MR intensity distribution compared to the analytical PDF and CL curves. In the left panel, the derived PDFs for $I_c/I_s = 0.1$ (thick line), $I_c/I_s = 1$ (thick dot-dashed line), and $I_c/I_s = 10$ (thick dashed line) are shown. The three thin lines are the expected PDFs for the three same cases. The curves for the right panel are the same. The three thin dotted lines in the right panel are the extrapolated CL curves. \label{f3b}}
\end{figure}

Simulated noise images using specific PDFs are used to test the algorithm in recovering the proper $1-3\times 10^{-7}$ detection threshold for known PDFs. To test the effect of the image area size on the CL extrapolation accuracy, images of various sizes are produced following a MR of $I_c/I_s$ equal to 0.1, 1 and 10 (see Fig.~\ref{f2}). For each size, 25 independent realizations are computed to derive the extrapolation accuracy. Fig.~\ref{f3b} and Table~\ref{tab1} show the CL extrapolation accuracy for simulated statistical distributions. In general, the algorithm slightly underestimates the $1-3\times 10^{-7}$ detection threshold for exponential statistics by $\sim$ 5\%, but usually within the 2$\sigma$ error calculated for each area size. Typically, the bigger the area is, the more accurate is the detection threshold. To achieve a detection threshold accuracy of 10\% for a $1-3\times 10^{-7}$ CL detection threshold, each PDF needs to be known up to a $1-10^{-4}$ CL (10,000 resolution elements per area). For instruments with smaller FOVs (several hundred per several hundred $\lambda/D$), an area of $\sim 50\times 50 \lambda/D$ (2,500 resolution elements) would deliver a detection threshold accuracy of $\sim $15\% for all types of distribution. A solution to increase the detection accuracy of small FOVs would be to combine observations of several objects of similar magnitudes and observing conditions to increase the number of independant noise realizations in each area.

\subsection{Simulated PSFs\label{simdata}}
The algorithm is now tested using simulated aberrated PSFs. For PSF observations, determination of the PDF is more complex. The speckle noise amplitude is decreasing with angular separation and the PDF may change with angular separation due to relative importance of random atmospheric speckles, photon, background and read noises. Since the speckle noise amplitude decreases with angular separation, a signal-to-noise ratio image is first obtained by dividing the pixel intensities, at each radius, by the standard deviation $\sigma$ of the noise at that radius (estimated using the IDL {\em robust\_sigma} algorithm). Finally, since CL are extrapolated, a compromise needs to be found between having a good radial sampling of the PDF and having sections of images big enough to adequately determine the PDF.

PSF simulations are performed using Fast Fourier Transforms of complex $2048\times 2048$ pixel images with a 512 pixels diameter pupil; the full width half maximum (FWHM) of the PSF is 4 pixels. The pupil has uniform amplitude and includes $\lambda/160$ RMS of phase errors generated using a power-law of index $-2.6$. The PSF images are then trimmed to $1024\times 1024$ pixels to avoid FFT aliasing effects. A non-aberrated reference PSF is subtracted to remove the Airy pattern and a signal-to-noise image is calculated. 

\begin{figure}
\epsscale{1}
\plotone{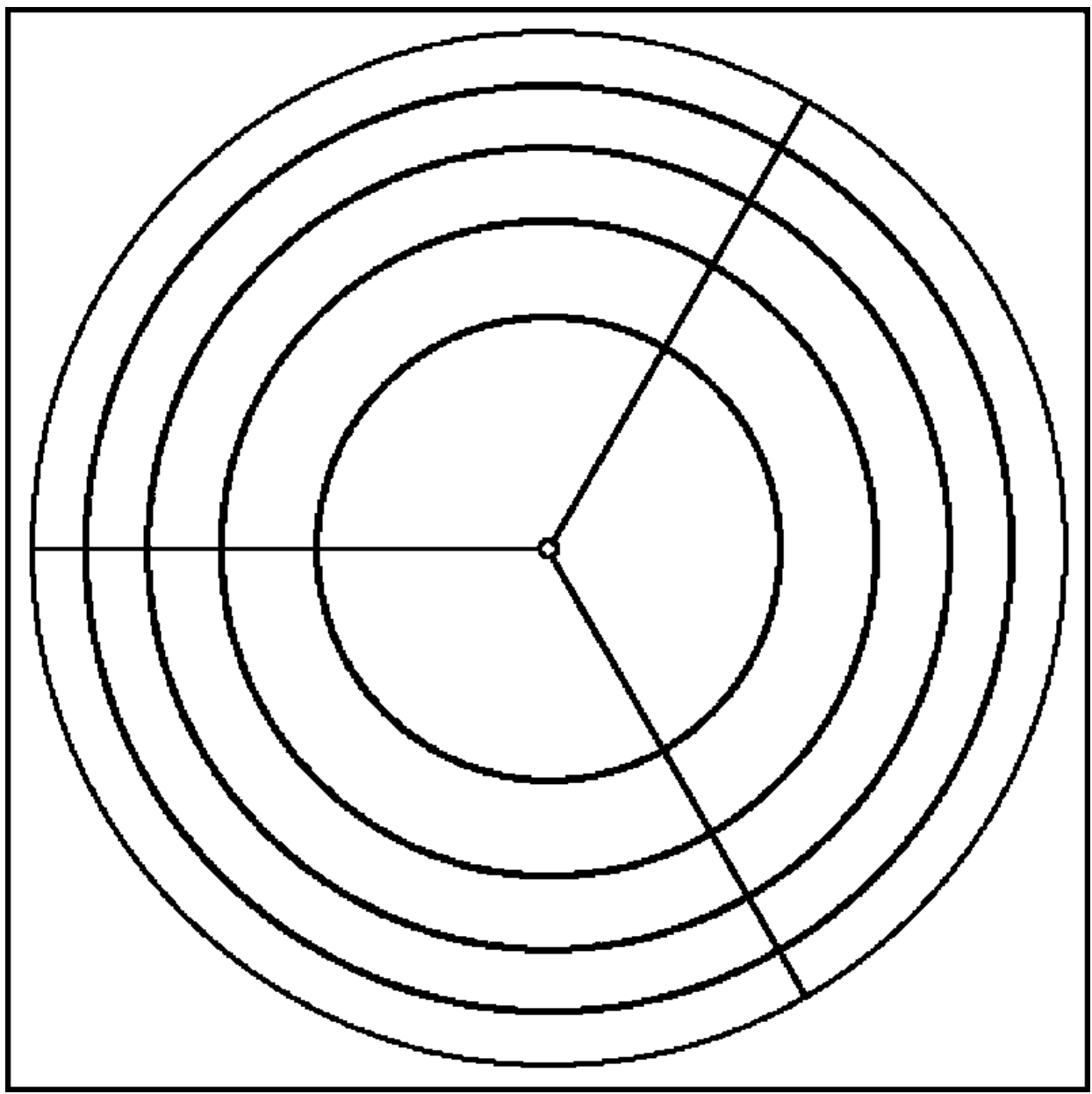}
\caption{Selected grid to calculate PDFs.\label{f3}}
\end{figure}

For simplicity, consider the calculation of the PDF within an annulus centered on the PSF. Since the presence of background or companion point sources inside that annulus could bias the statistics for real data (we are assuming that the background star density is such that only one or a few background objects are detected in the field of view (FOV) around any single target; cases with a high background star density will be discussed in section~\ref{hbsd}), we have chosen to divide the annulus in three azimuthal sections containing 50,000 pixels each ($\sim$~10,000 resolution elements, see Fig.~\ref{f3}). The median PDF over the three azimuthal sections is calculated. Given the area of these sections, the PDF will be known down to a $\sim 1-10^{-4}$ CL for Gaussian statistics and the $1-3\times 10^{-7}$ detection threshold will be known to $\sim $10\% accuracy (see Tab.~\ref{tab1}).

To further avoid cases where a point source is located at the border of two sections, the entire procedure is repeated by rotating the sections by 30 and 60 degrees which respect to the PSF center, and the median PDF over the three orientations is finally obtained. This procedure is repeated at different angular separations.

\begin{figure}
\epsscale{1}
\plotone{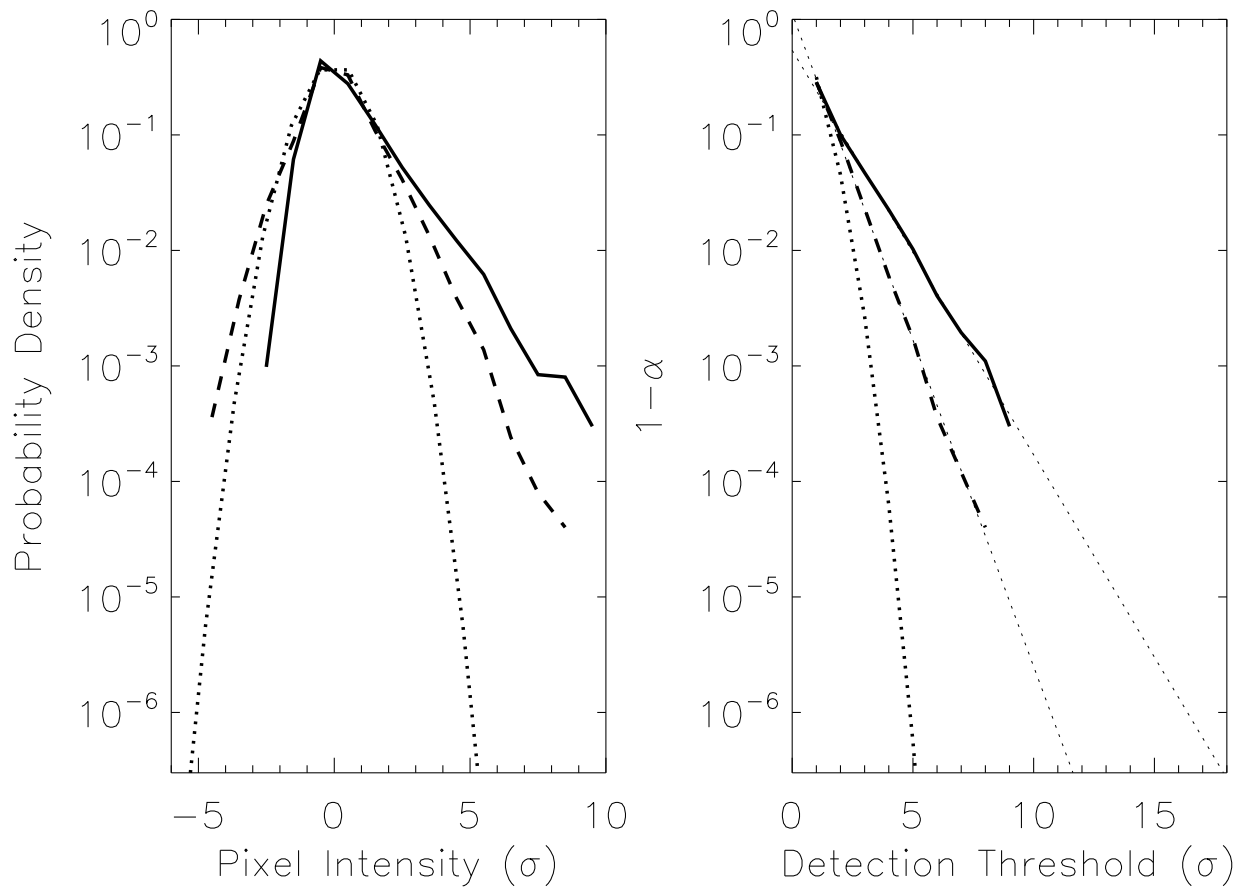}
\caption{PDFs and CLs for simulated PSFs. The dotted line is for a Gaussian noise distribution while the dashed and solid lines are respectively for simulated aberrated PSFs without/with a Gaussian pupil apodizer. The right panel shows the corresponding CL including the power-law fit (thin dotted lines) to estimate the $1-3\times 10^{-7}$ CL detection threshold.\label{f4}}
\end{figure}

\begin{figure}
\epsscale{1}
\plotone{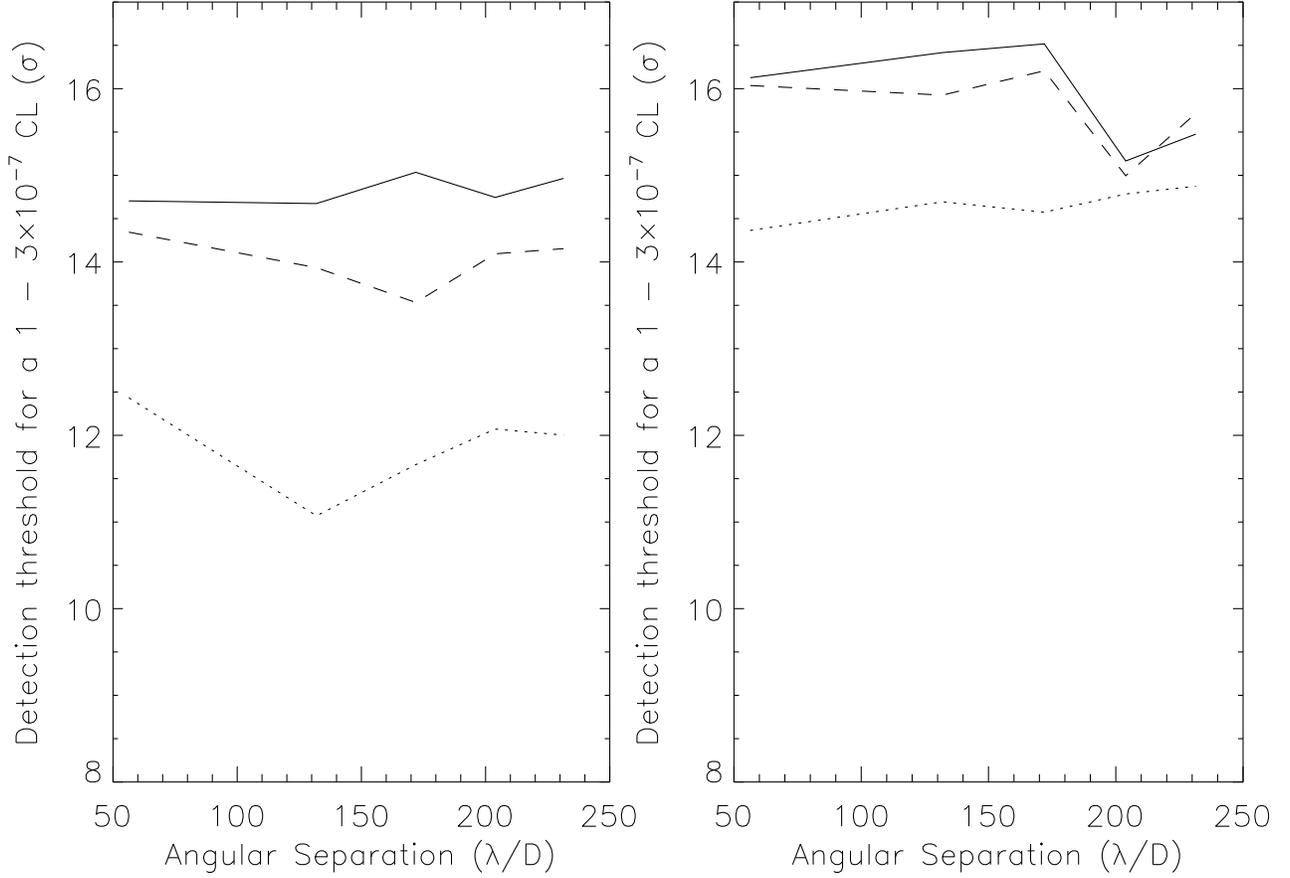}
\caption{Required detection threshold to obtain a $1-3\times 10^{-7}$ CL for simulated aberrated PSFs without (left) and with (right) a Gaussian pupil apodizer. The solid, dashed and dotted lines are for $\lambda/16$, $\lambda/32$ and $\lambda/160$ RMS of phase aberration (generated with a power-law of index $-2.6$), respectively.\label{f5}}
\end{figure}

Since the PDF is estimated in large areas that may contain speckles with $I_c \gg I_s$, $I_c \sim I_s$ or $I_c \ll I_s$, such technique returns an average PDF weighted by the various speckle noise contributions (pinned/unpinned speckles or Gaussian noises). Simulation with and without a coronagraph (simulated with a Gaussian pupil apodizer having a FWHM equal to a quarter of the pupil diameter) and for $\lambda/160$, $\lambda/32$ and $\lambda/16$ RMS phase aberration are presented (see Fig.~\ref{f4} and \ref{f5}). The algorithm clearly detects the MR distribution expected for a pinned speckle dominated PSF and the unpinned (exponential) speckle dominated coronagraphic PSF.

\begin{figure}
\epsscale{1}
\plotone{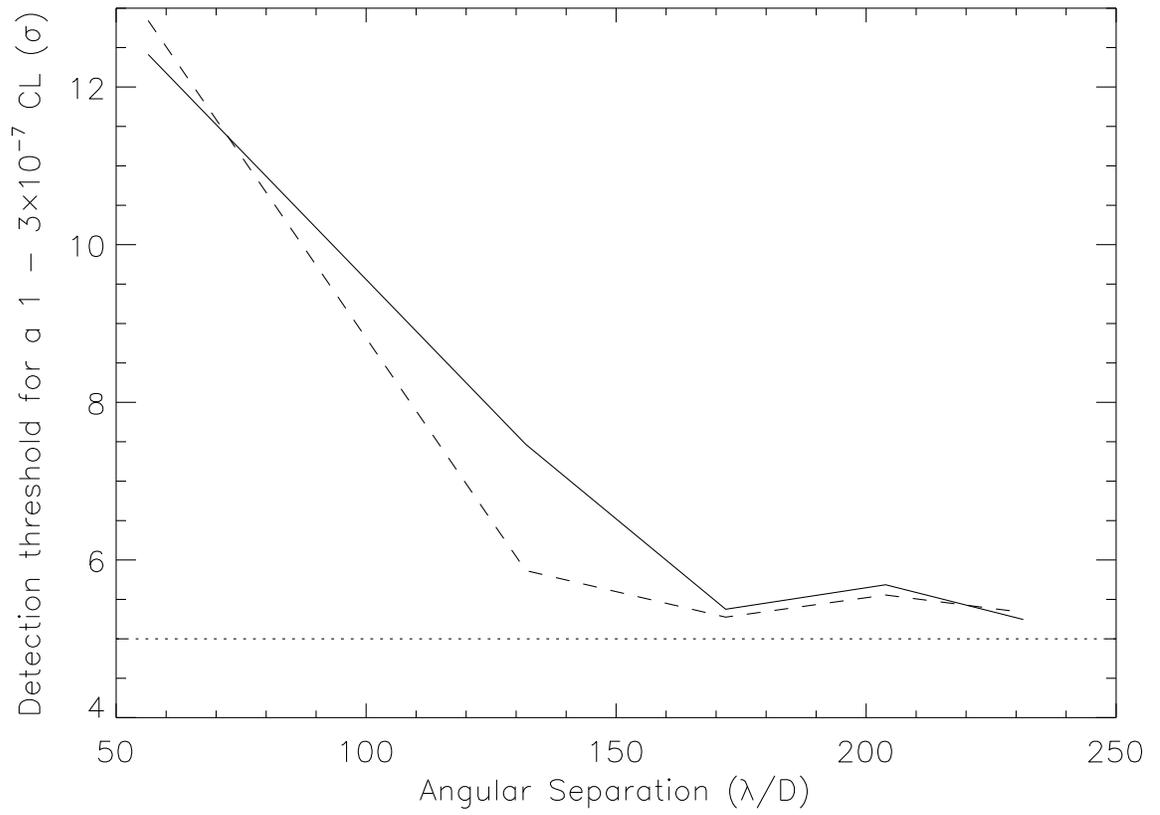}
\caption{See Fig.~\ref{f5}. A random Gaussian pixel-to-pixel noise is added to the PSF image such that it dominates at wide separations.\label{f6}}
\end{figure}

The non-coronagraphic $\lambda/160$ RMS simulations confirm that speckle noise follows a MR (required detection threshold of $\sim 10\sigma$ for a $1-3\times 10^{-7}$ CL), as expected since pinned speckles are dominant for this case. As the quantity of aberrations increases, the ratio of pinned to non-pinned speckles decreases and the noise becomes exponential. For the Gaussian apodized case, since pinned speckles are strongly attenuated, the halo term dominates and the noise is more exponential. Note that none of these curves are expected to be flat as a function of angular separation since the ratio of pinned to unpinned speckles is varying with angular separation, thus changing the pixel intensity distribution, and some noise is expected from the CL curve extrapolation (see Tab.~\ref{tab1}). Another simulation was performed using the $\lambda/160$ RMS case to show that if a constant Gaussian noise (background or read noise) is added to the image, the algorithm correctly detects the change of intensity distribution of the pixels at wide separations (see Fig.~\ref{f6}).

\begin{figure}
\epsscale{1}
\plotone{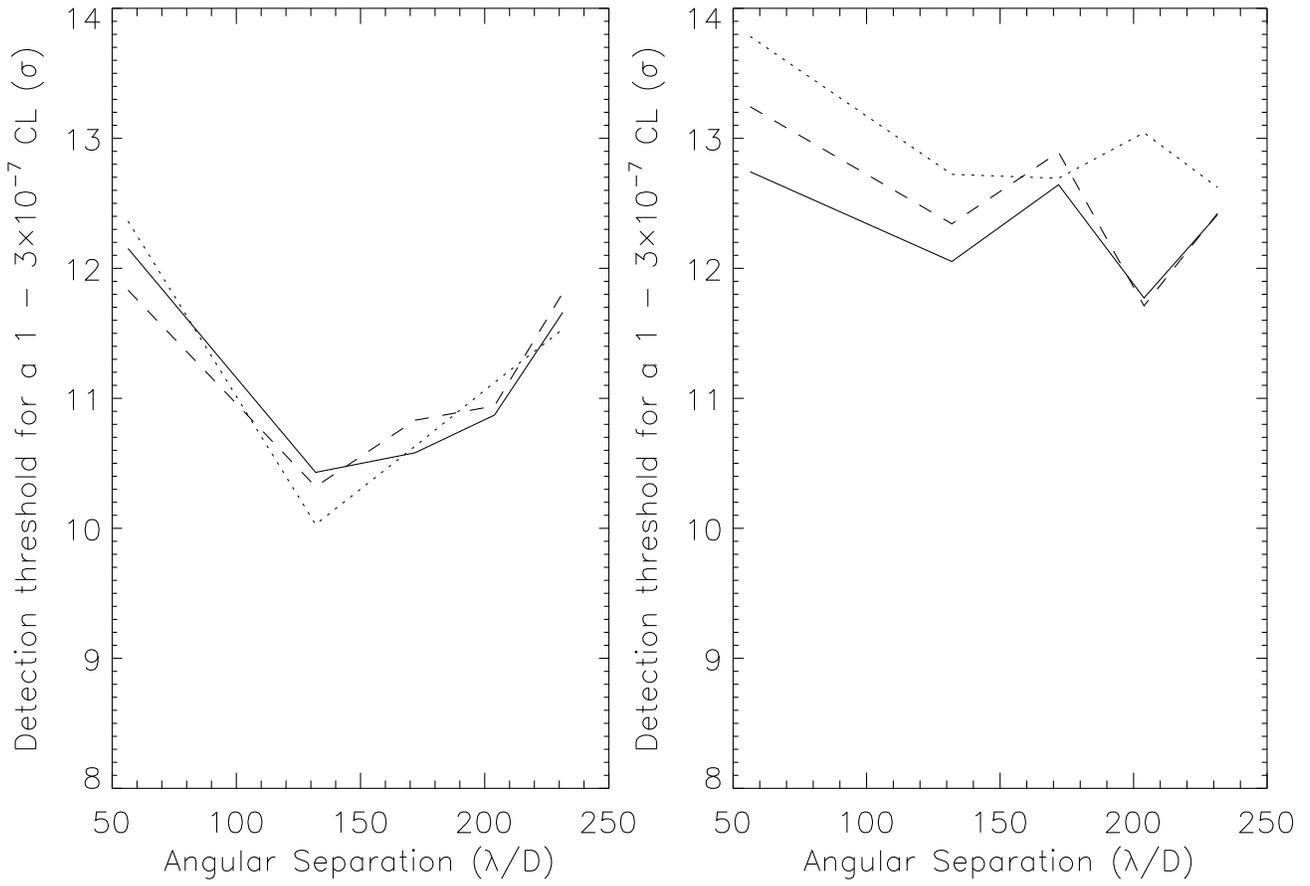}
\caption{See Fig.~\ref{f5}. A partially correlated reference PSF is subtracted (see text for details).\label{f7}}
\end{figure}

In high-contrast imaging observations, a partially correlated reference star PSF is usually subtracted to remove a fraction of the quasi-static speckle noise. Such reference PSF can be obtained by observing a nearby target, by acquiring the same star at another wavelength (simultaneous spectral differential imaging, \citealp{marois2005}) or polarization \citep{potter2001}, or by building the reference using images acquired with different field angles (angular differential imaging, \citealp{marois2006}). Such a PSF subtraction is now simulated to estimate how its affect the PDF. The observed PSF $I$ is simulated with a $\lambda/160$ RMS phase aberration $\phi$, with and without a Gaussian apodizer. The reference PSF $I_{\rm{ref}}$ is constructed by combining a perfectly correlated phase aberration $a\phi$, where $a$ is a constant less than 1, with an uncorrelated part $\Delta \phi$ such as:

\begin{equation}
I = | \rm{FT}(Ae^{i\phi})|^2
\end{equation}
\begin{equation}
I_{\rm{ref}} = | \rm{FT}(Ae^{i(a\phi+\Delta \phi)})|^2 \rm{,}
\end{equation}
\noindent where the total noise RMS of $\phi$ and $a\phi+\Delta \phi$ are equal and the ratio of the noise RMS of $a\phi$ and $\Delta \phi$ is equal to 0.1, 1 and 10. Unless the background, read noise, random atmospheric speckles or photon noises are achieved by the reference PSF subtraction, the residual PDF is essentially unchanged for cases with and without a coronagraph (see Fig.~\ref{f7}).

\section{Application to Observational Data\label{obsdata}}
\begin{figure}
\epsscale{0.55}
\plotone{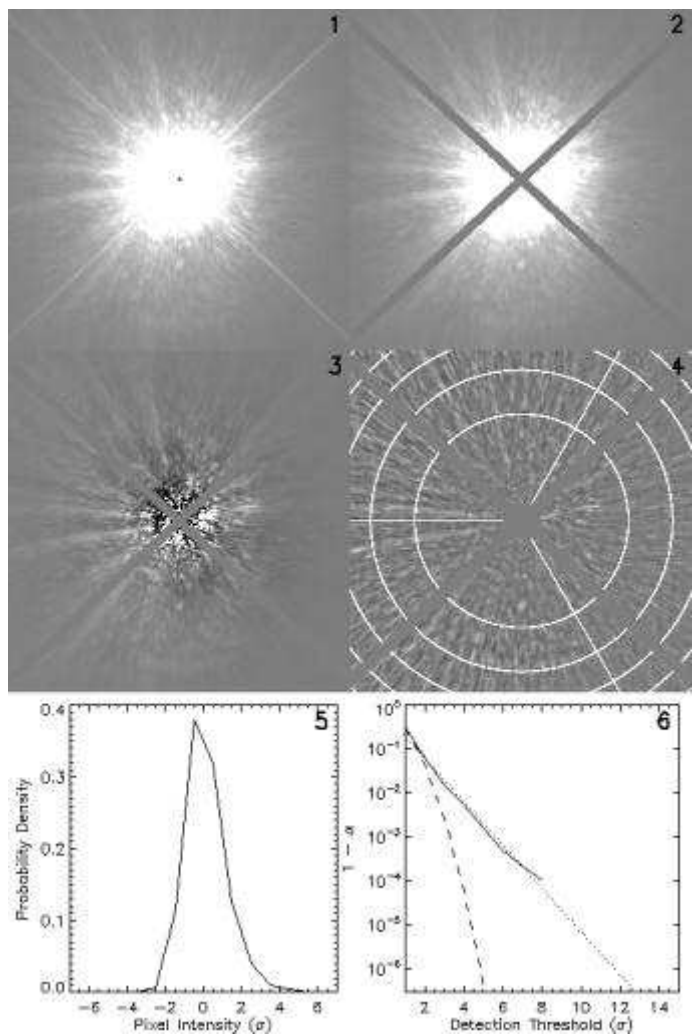}
\caption{The different steps involved in estimating a detection threshold in a specific region of the PSF using the confidence level approach. Panel 1 shows a Gemini CH4-short saturated PSF image that has been reduced and registered to the image center. Panels 2, 3 and 4 show respectively the same PSF image but with the secondary support structure diffraction masked, after noise filtering, and after noise normalizing. Panels 5 and 6 show respectively the pixel intensity distribution (PDF) and corresponding CL curve inside a typical region of the PSF shown in panel 4. An extrapolation of the CL curve gives the $1-3\times 10^{-7}$ confidence level detection threshold for that region (here approximately 12.6$\sigma$). Steps illustrated by panels 5 and 6 are repeated for all regions of the PSF and after rotating the regions by 30 and 60 degrees to eliminate the bias resulting from point sources located at the edge of two regions.\label{figexample}}
\end{figure}

The steps required to use the algorithm with observational data are similar to the ones described in Tab.~\ref{tabpdfstep} and $\S~\ref{simdata}$. Only a few additional reduction steps are necessary. Fig.~\ref{figexample} illustrates the various steps of the technique using Gemini data.

Beside the usual data reduction, deviant pixels, like diffraction from the secondary mirror support, must first be masked. Diffraction from the secondary mirror support usually produces a bright concentrated flux emanating from the PSF core along several azimuthal directions. Since this flux is not produced by quasi-static aberrations and is very localized in the image, if we include these pixels in the PDF, they will produced a bright positive tail in the PDF and the $1-3\times 10^{-7}$ CL detection threshold will be overestimated.

Since the main science goal is to detect point sources, noise filtering is also applied to remove the noise that is not at the spatial scale of point sources. An $8\times 8$ FWHM median filter is first subtracted from the image to reject large spatial period noises. Then, a $1\times 1$ FWHM median filter is applied to reject bad/hot pixels and smooth out the noise having spatial period below the resolution limit. The image is finally divided, at each radius, by the standard deviation $\sigma$ of the noise (again obtained with the IDL {\em robust\_sigma} algorithm) to obtain a signal-to-noise ratio image.

\begin{figure}
\epsscale{1}
\plotone{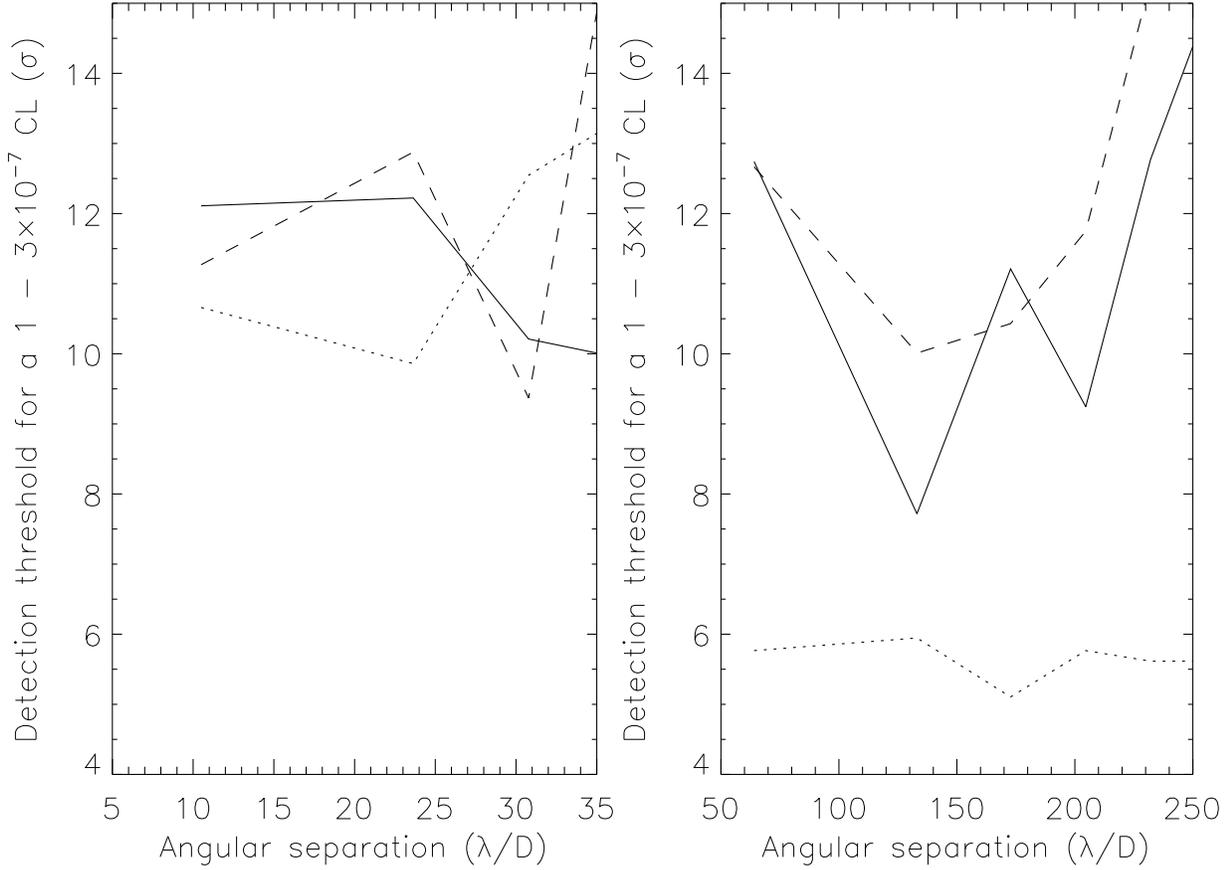}
\caption{Left panel: TRIDENT SSDI required detection threshold for a $1-3\times 10^{-7}$ CL of the star Ups And acquired at the Canada-France-Hawaii telescope. Solid, dashed and dotted lines show the required detection limit for respectively the TRIDENT PSF, the SSDI simple difference subtraction (subtraction of two images acquired simultaneously at two wavelengths) and the SSDI simple difference minus the reference SSDI simple difference image of a reference star. Right panel: same plot but for the ADI technique and the star HD97334B acquired at Gemini. Solid, dashed and dotted lines are for respectively a single PSF image minus a symmetric radial profile (spiders masked), a single ADI reference subtracted image and the total ADI median combined image (median of 90 ADI-subtracted images).\label{f8}}
\end{figure}

The algorithm is first tested using data obtained at the Gemini telescope with the Altair adaptive optics system \citep{saddlemyer1998} and the NIRI near-infrared camera \citep{hodapp2000}. These data are part of the Gemini Deep Planet Survey \citep{lafreniere2007a} that uses the angular differential imaging (ADI) technique \citep{marois2006,lafreniere2007b} to detect faint companions. This technique consists in acquiring a sequence of images with continuous FOV rotation. A reference PSF that does not contain any point sources is first obtained by combining images of the sequence, and the quasi-static speckle noise is then attenuated by subtracting the reference ADI PSF. The data for the star HD97334B (program GN-2005A-Q16), acquired on April 18, 2005 with good seeing conditions (Strehl of $0.2$ at H-band), are presented. These data have been reduced, registered and processed using the pipeline described in \citet{marois2006} with the additional steps mentioned above, i.e. pixel masking and noise filtering \& normalizing. Given that NIRI images are $1024\times 1024$ pixels and PSFs have 3 pixels per FWHM, we have chosen the same areas as the simulated PSFs mentioned above (see \S~\ref{simdata}) to calculate the PDF. For each region of the image, the pixel intensity histogram is obtained and then integrated to derive the CL curve. The CL is then extrapolated using a polynomial fit and the $1-3\times 10^{-7}$ CL detection threshold is estimated. The derived detection thresholds for a $1-3\times 10^{-7}$ CL are presented (see Fig.~\ref{f8}) for a single PSF, a PSF minus the ADI reference PSF, and the combined ADI-subtracted images.

The algorithm is next tested using observational data obtained at the Canada-France-Hawaii telescope using the PUEO adaptive optics system \citep{rigaut1998} and TRIDENT near-infrared camera \citep{marois2005}. TRIDENT is a triple beam multi-wavelength (1.58, 1.625 and 1.68$\micron$ with 1\% bandpass) imager built following the simultaneous spectral differential imaging technique \citep{racine1999,marois2000}. The technique consists in acquiring several images at different wavelengths and subtracting them to attenuate the speckle noise while retaining most of the flux of nearby companions. These data have been acquired as part of a direct imaging survey of stars confirmed to possess exoplanets from radial velocity analysis. The dataset of the star Ups And, acquired on November 14, 2002, is used here. Seeing conditions were relatively good and the Strehl ratio was of the order of $0.5$. The data have been reduced by subtracting a dark, dividing by a flat field, correcting for bad/hot pixels, and registering the PSF at the image center. An optimized reference PSF was obtained using a star (Chi And) having a similar spectral type, magnitude and acquired at the same DEC and HA to minimize PSFs evolution from differential atmospheric refraction and flexure effects. Performances with and without the simultaneous reference PSF and the Chi And reference PSF subtractions are analyzed. Due to the limited FOV of TRIDENT, sections of 10,000 pixels (500 resolution elements, given that TRIDENT has 5 pixels per $\lambda/D$) are used to derive the PDF. Detection thresholds are thus known to $\sim$15\% (see Tab.~\ref{tab1}).

\begin{figure}
\epsscale{1}
\plotone{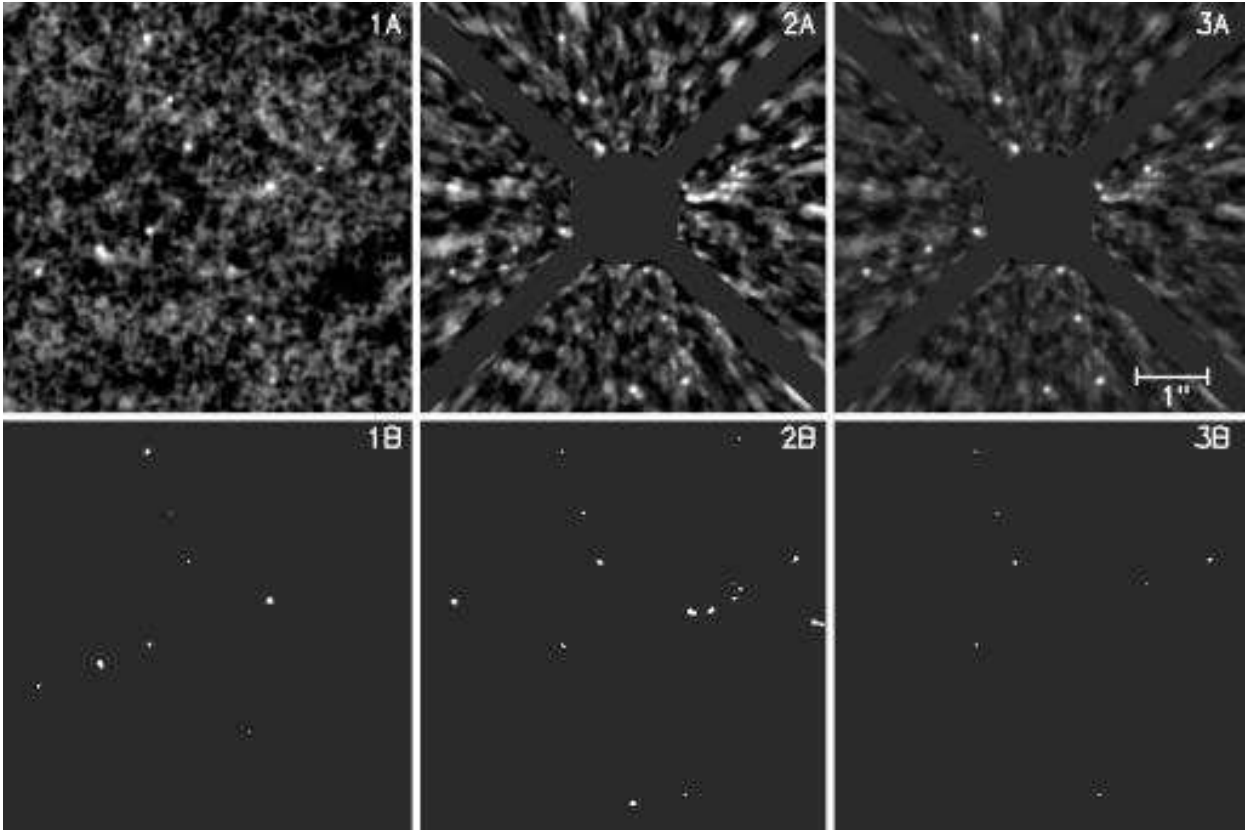}
\caption{$1-3\times 10^{-7}$ confidence level detections for a Gaussian distributed noise and Gemini data. After analyzing the noise statistical distribution, artificial point sources are added at 0.9, 1.6 and 2.5$^{\prime \prime}$ from the image center at PA of 20, 110, 200 and 290 degrees from the vertical axis and each having an intensity equal to $5\sigma$ of the noise. Panel 1A shows the image generated using a Gaussian noise (display with a linear intensity range between -1$\sigma$ and 5$\sigma$). Panel 1B shows the pixels that are higher than 5$\sigma$. Panels 2A and 2B are similar but for the Gemini PSF. Panels 3A and 3B are again similar but for artificial point sources now having an intensity equal to 10$\sigma$ of the noise, the detection threshold derived using the CL technique. FOV is $5.65^{\prime \prime} \times 5.65^{\prime \prime}$. Panel 3A is displayed with a linear intensity range between -2$\sigma$ and 10$\sigma$. \label{figexample2}}
\end{figure}

It is clear that both TRIDENT and Gemini raw PSFs are limited by a non-Gaussian noise. For both TRIDENT/CFHT and NIRI/Gemini images, even after subtraction of a reference PSF, the residuals are still dominated by quasi-static speckles rather than averaged atmospheric speckle, background, read, or photon noises. Only the final combined ADI-subtracted image possessed a clear Gaussian-like noise. Fig.~\ref{figexample2} shows a visual example of a 5$\sigma$ detection with and without a Gaussian distributed noise after introducing artificial 5$\sigma$ point sources. For the Gaussian distributed noise, only the artificial point sources are detected with a detection threshold at 5$\sigma$,\footnote{Only approximately half of the artificial point sources are detected $\ge 5\sigma$ since the artificial sources, being $5\sigma$ in intensity, vary in S/N by $1\sigma$ RMS due to the underlying noise in the image. A $5\sigma$ detection threshold thus misses/detects $\sim 50$\% of $5\sigma$ sources.} while for the Gemini data (a MR distributed noise), numerous false positive sources are observed for the same detection threshold. If instead we select the $1-3\times 10^{-7}$ CL detection threshold obtained by the technique described in this paper (here equal to $10\sigma$, see Fig.~\ref{f8}), then only the artificial point sources are detected. It is interesting to note that the artificial source detection CL in the left and right panels of Fig.~\ref{figexample2} are the same.

\section{Discussions\label{disc}}

\subsection{PDF Evolution with Quasi-Static Speckle Averaging\label{dis}}
It was shown in \S~\ref{obsdata} that the ADI technique produces a quasi-Gaussian noise. This is mainly due to the FOV rotation that occurs during the observing sequence; the residual noise is averaged incoherently when combining the images after FOV alignment. From the central limit theorem, it is thus expected that the noise in the final combined image shows quasi-Gaussian statistics.

\begin{figure}
\epsscale{1}
\plotone{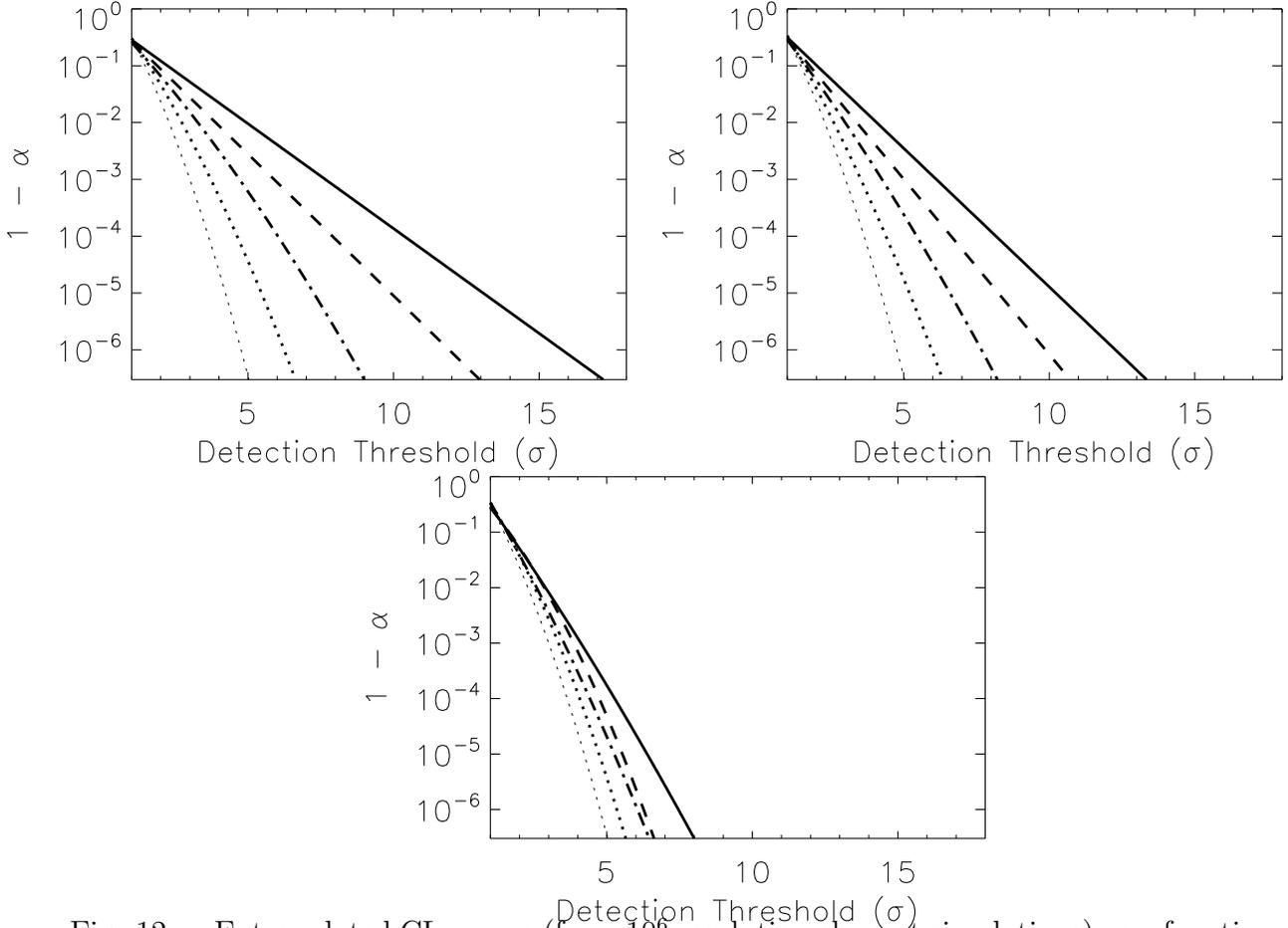}
\caption{Extrapolated CL curves (from 10$^{6}$ resolution element simulations) as a function of the detection threshold for various $n_{\rm{eff}}$. Upper left panel is for an MR statistical distribution with $I_c/I_s = 0.1$ while the upper right and bottom panels are for $I_c/I_s = 1$ and $I_c/I_s = 10$ respectively. The solid thick line is for a single noise realization while the thick dashed, dot-dashed and dotted lines are for the average of 2, 5 and 25 independent noise realizations. The thin dotted line is the CL curve for a Gaussian noise intensity distribution.\label{f9}}
\end{figure}

\begin{figure}
\epsscale{1}
\plotone{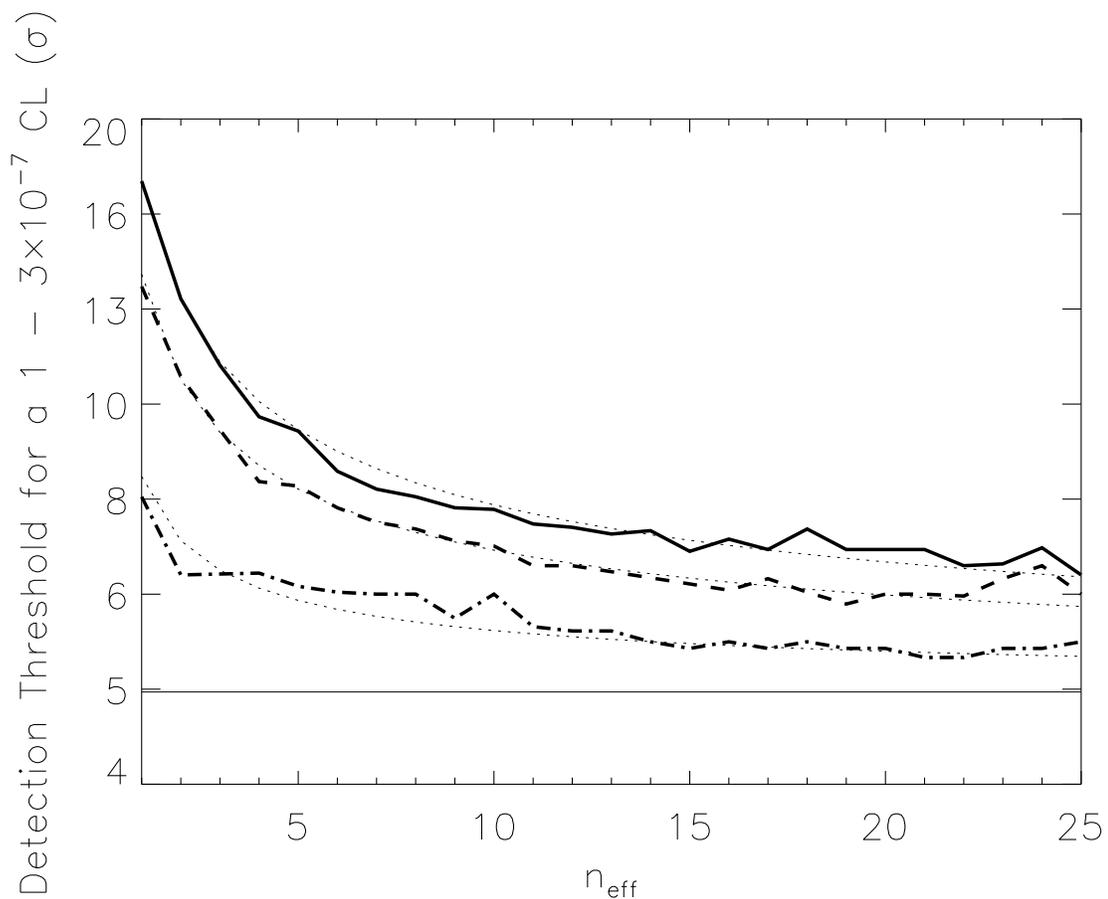}
\caption{The $1-3\times 10^{-7}$ CL detection threshold as a function of $n_{\rm{eff}}$, the number of independent noise realizations. The thick line is for an MR statistical distribution with $I_c/I_s = 0.1$ while the thick dashed and dot-dashed lines are for $I_c/I_s = 1$ and $I_c/I_s = 10$ respectively. The thin solid line is for Gaussian statistics. The three thin dotted lines are the predicted detection thresholds derived from the fitted power-law (see text for details).\label{f10}}
\end{figure}

In this section, simulations are presented to estimate the number of independent speckle noise realizations required to converge to a quasi-Gaussian noise intensity distribution. Random noise images having 10$^{6}$ resolution elements are created following an MR distribution having $I_c/I_s$ equal to 0.1, 1 and 10. The PDF and CL curves are calculated for a single realization up to the coaddition of 25 independent realizations (see Fig.~\ref{f9}). Typically, $\sim $20 independent realizations are required for the MR distribution to converge, to $\sim $20\%, to a Gaussian distribution. Fig.~\ref{f10} shows the detection threshold $d$ for a $1-3\times 10^{-7}$ CL as a function of $n_{\rm{eff}}$, the number of independent noise realizations
\begin{equation}
n_{\rm{eff}} = n \frac{t_{\rm{exp}}}{\tau_{\rm{dcorr}}} \rm{,}\label{eqneff}
\end{equation}
where $n$ is the number of acquired images in the sequence, $t_{\rm{exp}}$ is the integration time per image and $\tau_{\rm{dcorr}}$ the speckle noise decorrelation timescale (the equation is valid if $\tau_{\rm{dcorr}} \ge t_{\rm{exp}}$; if $\tau_{\rm{dcorr}} < t_{\rm{exp}}$ then $n_{\rm{eff}} = n$). These three curves can be well fit by a simple power-law of the form
\begin{equation}
d(n) = [d_1 - 5] n_{\rm{eff}}^{-0.63} + 5 \rm{,}\label{eqevol}
\end{equation}
\noindent where $d_1$ is the detection threshold of a single image for a $1-3\times 10^{-7}$ CL. This equation is valid for all types of statistical distributions studied here. Eq.~\ref{eqevol} can be used to predict the detection threshold required for a $1-3\times 10^{-7}$ CL and a statistical distribution with a known instantaneous PDF and speckle noise decorrelation timescale. If we consider the Gemini ADI observation ($d_1 \sim 13$ for a single ADI-subtracted image, see Fig.~\ref{f8}), for a 70 minute observation sequence with $\tau_{\rm{dcorr}} \sim 1.5$ minutes (at 2$^{\prime \prime}$ or 50$\lambda/D$, see \citealp{marois2006}), it is expected that the detection threshold for a $1-3\times 10^{-7}$ CL of the final combined image ($n_{\rm{eff}} = 46.7$) will be $\sim 5.7$ at 50$\lambda/D$, in good agreement with the number derived with real images ($\sim 5.8$, see Fig.~\ref{f8}).

\subsection{Targets with a High Background Star Density\label{hbsd}}
In some cases, it is desirable to observe an interesting nearby target situated along the galactic plane or in front of the galactic bulge. In those areas, the high background stellar density implies that numerous background stars will be present in each of the areas defined to derive pixel intensity distributions. The detection threshold obtained will be affected by those stars since the algorithm would consider them as speckles. Several techniques can be used to remove the stars before estimating detection thresholds. A simple solution is to subtract these stars, using a non-saturated image of the primary, and mask any remaining contaminated areas with a not-a-number (NaN) mask. If the observations are obtained using the ADI technique then a star-free residual image can be obtained in the following manner. Prior to combining all the ADI-subtracted images together, instead of rotating them by the angle required to align their field of view, they are rotated by the negative of that angle such that all off-axis sources (the background stars) are eliminated by the median combination. As the amplitude of the rotation between the images is the same as for the ``proper'' combination, the effect of the median combination on the residual noise is expected to be the same and this star-free residual image can be used to estimate the noise distribution. Of course, the proper ADI combination of images must be used to search for companions. Another approach, still within the ADI framework, is to use the final ADI residual image to subtract the off-axis sources from each non-rotated ADI-subtracted image. Then these source-free images are rotated by the negative of the angle needed to align their field of view, such that their median combination eliminates the off-axis sources subtraction residuals. As for the previous technique, this source-free residual image should have the same residual noise distribution as the proper ADI residual image.

\section{Conclusion\label{con}}
A robust technique was elaborated to estimate sensitivity limits using a CL approach. This technique correctly finds the expected MR intensity distributions of simulated and real PSFs, and properly detects a change of PDF as a function of angular separation. Experiments with simulated and observational data confirm the prediction of the theory that raw PSFs obtained with high-contrast imaging instruments are limited by a non-Gaussian noise. A correction factor (up to 3) needs to be applied to detection limits found assuming Gaussian statistics to obtain the desired $1-3\times 10^{-7}$ CL detection threshold. Properly estimating this effect is important for future high-contrast imaging instruments for both ground- and space-based dedicated missions since a loss of a factor of three in contrast results in less sensitivity to low-mass exoplanets or, if a specific contrast needs to be achieved, integration times need to be at least nine times longer. It was shown that the ADI technique is the only observing strategy currently known that generates, intrinsically, a quasi-Gaussian noise at all separations where sufficient FOV rotation has occurred. A simulation has shown that it takes typically $\sim 20$ independent speckle noise realizations to produce an average speckle noise that shows quasi-Gaussian statistics. A general power-law is derived to predict the detection threshold required when averaging independent speckle noise realization of known PDFs and decorrelation timescale.

\acknowledgments
The authors would like to thanks R\'{e}mi Soummer, Mike Fitzgerald, James Graham, Anand Sivaramakrisnan, Lisa Poyneer and Daniel Nadeau for discussions. This research was performed under the auspices of the US Department of Energy by the University of California, Lawrence Livermore National Laboratory under contract W-7405-ENG-48, and also supported in part by the National Science Foundation Science and Technology Center for Adaptive Optics, managed by the University of California at Santa Cruz under cooperative agreement AST 98-76783. This work is also supported in part through grants from the Natural Sciences and Engineering Research Council, Canada and from the Fonds Qu\'{e}b\'{e}cois de la Recherche sur la Nature et les Technologies, Qu\'{e}bec.

\clearpage
\begin{table}
\begin{center}
\caption{Basic steps to derive the PDF, to estimate the CL curve and to determine the $1-3\times 10^{-7}$ CL detection threshold.\label{tabpdfstep}}
\begin{tabular}{cc}\hline
Step & Action \\ \hline \hline
1 & Define an image region\\
2 & Subtract the mean intensity\\
3 & Divide the pixel intensities by the noise RMS value\\
4 & Obtain the pixel intensity histogram (PDF)\\
5 & Integrate PDF between $\pm$d to obtain the corresponding CL curve\\
6 & Perform a polynomial fit on the CL curve\\
7 & Extrapolate the CL curve to derive the $1-3\times 10^{-7}$ CL detection threshold\\
\hline
\end{tabular}
\end{center}
\end{table}

\clearpage
\begin{table}
\begin{center}
\caption{The $1-3\times 10^{-7}$ CL detection threshold $d$ derived by extrapolating different type of MR statistics as a function of the number of resolution elements. MR01 is a modified Rician distribution with $I_c/I_s = 0.1$ while MR1 and MR10 are for $I_c/I_s = 1$ and $I_c/I_s = 10$ respectively. Both the mean and standard deviation of the detection threshold are obtained by analyzing 25 independent noise realizations.\label{tab1}}
\begin{tabular}{ccccc}\hline
Statistics & Number of  & Expected $d$  & $<d>$ & Stddev($d$)\\
 & res. element & ($\sigma$) & ($\sigma$) & ($\sigma$) \\ \hline \hline
Gaussian & 10$^{3}$ & 5.0 & 5.4 & 1.1 \\
 & 10$^{4}$ &  & 5.33 & 0.40 \\
 & 10$^{5}$ & & 5.06 & 0.29 \\
 & 10$^{6}$ & & 5.06 & 0.11 \\ \hline
MR10 & 10$^{3}$ & 7.7 & 9.3 & 1.9 \\
 & 10$^{4}$ & & 8.20 & 0.95 \\
 & 10$^{5}$ & & 8.02 & 0.59 \\
 & 10$^{6}$ & & 8.02 & 0.66\\ \hline
MR1 & 10$^{3}$ & 13.5 & 14.9 & 2.5 \\
 & 10$^{4}$ & & 13.95 & 0.98 \\
 & 10$^{5}$ & & 13.44 & 0.50 \\
 & 10$^{6}$ & & 13.48 & 0.37\\ \hline
MR01 & 10$^{3}$ & 18.2 & 18.7 & 2.6 \\
 & 10$^{4}$ & & 17.2 & 1.4 \\
 & 10$^{5}$ & & 17.13 & 0.40 \\
 & 10$^{6}$ & & 17.18 & 0.41 \\
\hline
\end{tabular}
\end{center}
\end{table}

\clearpage

\clearpage

\clearpage

\clearpage

\clearpage

\clearpage

\clearpage

\clearpage

\clearpage

\clearpage

\clearpage

\clearpage

\clearpage
\end{document}